\documentclass[11pt]{article}
\usepackage{graphicx,moriond,epsfig}

\begin{document}
\vspace*{4cm}
\title{NEARBY STAR CLUSTER FORMATION: PROBES OF COSMOLOGICAL GALAXY EVOLUTION}

\author{P. ANDERS (1), R. DE GRIJS (2), U. FRITZE-V.ALVENSLEBEN (1)}

\address{(1) Institut f\"ur Astrophysik, Friedrich-Hund-Platz 1, 37077
G\"ottingen, Germany; (2) Department of Physics \& Astronomy, University of
Sheffield, Hicks Building, Hounsfield Road, Sheffield, S3 7RH, UK}

\maketitle

\abstracts{The old Star Cluster (SC) systems surrounding any sofar
investigated  galaxy represent a powerful tool for the understanding of the 
cosmological evolution of their host galaxies. Phases of enhanced cluster
formation can be identified by reliably age  dating the clusters. These are
symptoms of violent star formation,  results of  intense starbursts as
commonly seen in merging nearby  galaxies and high-z  forming galaxies. 
However, the young SC systems of nearby merging galaxies and the  old SC 
systems of nearby passive galaxies (most likely resulting from high-z 
starburst events of these galaxies) appear to be significantly different  in
terms of their mass functions. Whether this difference originates in  
differences in the formation physics/environment or is only due to 
dynamical evolution of the cluster systems is currently a hot issue.  A main
diagnostic  for the survival probability of a newly formed cluster  is most
likely its compactness, since less compact, fluffy clusters are  more
vulnerable to destructive interactions  with the galactic environment  of
the cluster.
For these reasons, we developed 
1) a tool to significantly improve photometry of nearby SCs (which appear 
resolved on high-resolution images, provided e.g. by {\sl HST}), by
accurately  measuring and taking the cluster size into account, and 
2) a tool to effectively determine ages and masses (plus metallicities and 
extinctions) of SCs from multi-wavelength  photometry, ranging from the UV 
to the NIR.
We will present these tools and applications to SC systems in different 
nearby SC-forming galaxies, comprising a wide range of environments for the 
newly born clusters. These applications, though preliminary, will help to 
disentangle the effects leading to the observed differences in SC systems.}

\section{Why young nearby star clusters?}

\begin{figure}
\begin{center}
\psfig{figure=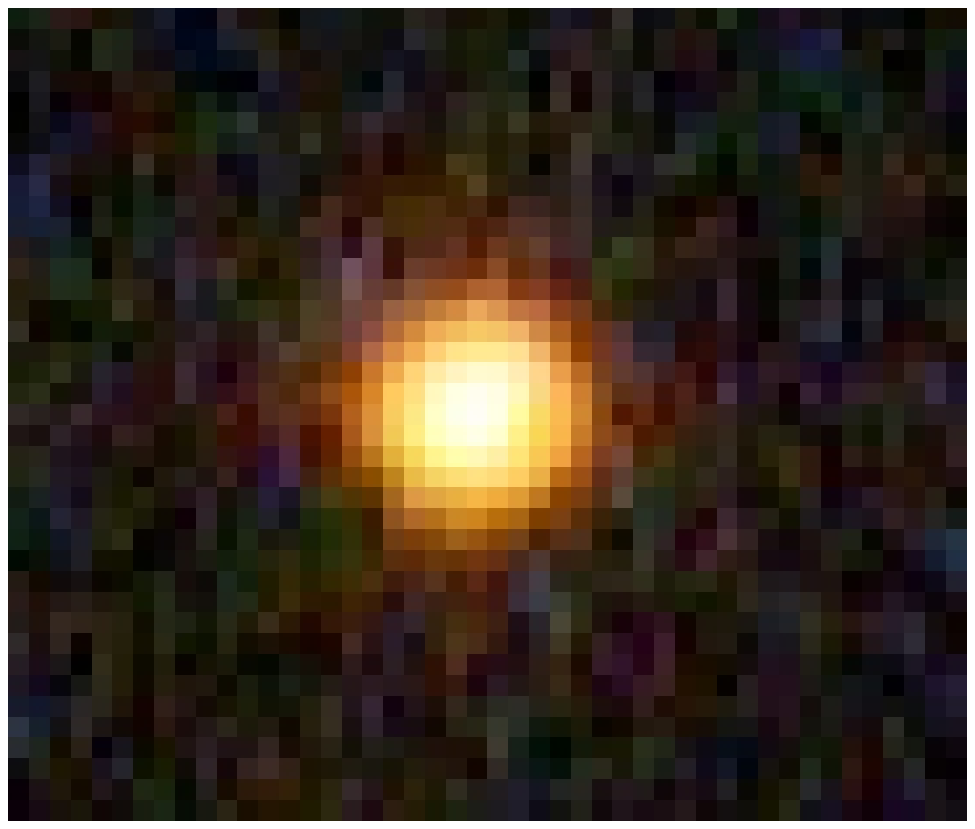,height=1.7in}
\psfig{figure=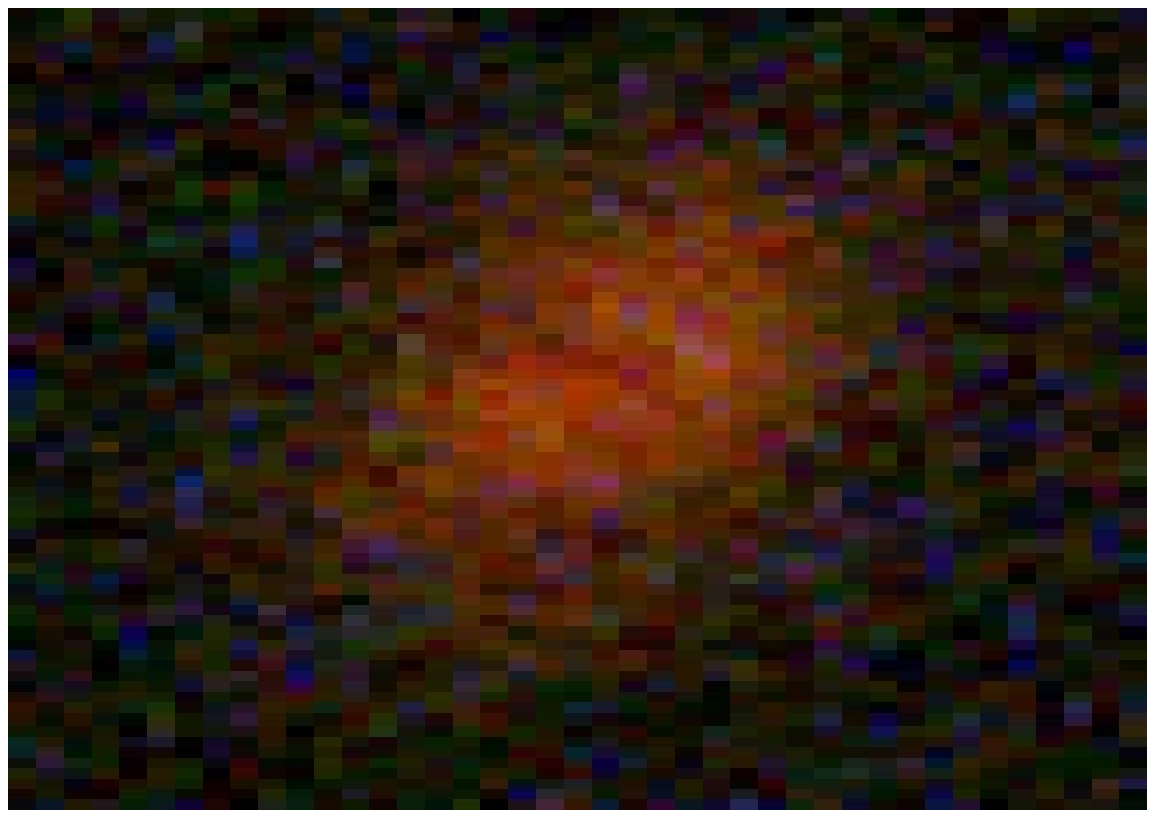,height=1.7in}
\psfig{figure=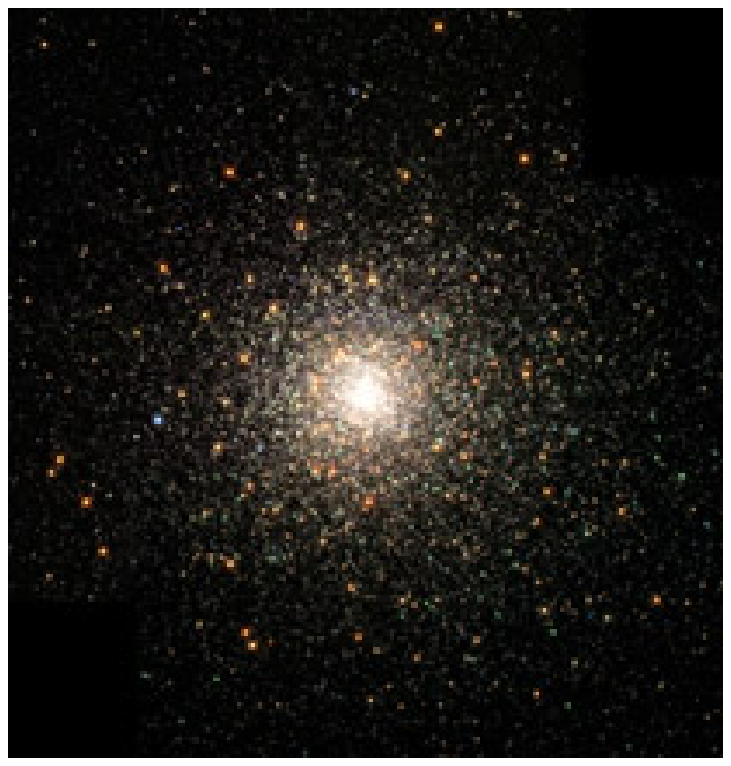,height=1.7in}
\caption{Left \& middle panel: 2 distant starburst galaxies, zoom-ins of
the {\sl HST} UDF. Right panel: Milky Way globular cluster M80.}
\label{fig:sb}
\end{center}
\end{figure}

Strongly bound, hence long-lived star clusters present a fossil record of
the violent star formation history of their host galaxies. (Almost) all
galaxies known contain old star clusters, globular clusters, with ages
around a Hubble time. The formation conditions of this cluster population
give vital clues to the conditions during the formation of the host galaxy
itself. However, due to the high redshift of galaxy formation (and
accompanying globular cluster formation) and the related limited spatial
resolution (see Fig. \ref{fig:sb}, left \& middle panel) this process cannot be studied
directly. Globular clusters are characterised by a tidally truncated light
profile (King 1962), they are relaxed, round systems (see Fig. \ref{fig:sb}, 
right panel), and the GC system as a whole shows a Gaussian-shaped luminosity/mass 
function.

\begin{figure}
\begin{center}
\psfig{figure=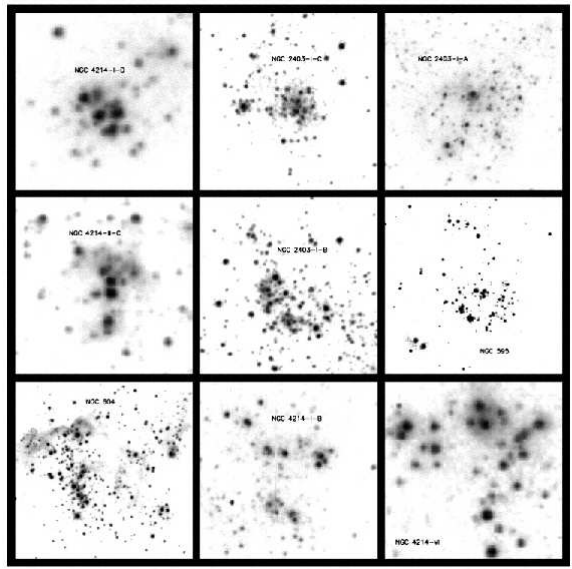,height=3in}
\caption{{\sl HST} U-band images of several SOBAs. Host galaxy indicated in the
plots. From Maiz-Apellaniz (2001).}
\label{fig:maiz}
\end{center}
\end{figure}

On the other hand, young clusters with a large range of dynamical, 
structural and photometric properties are currently produced in all
starburst environments (for a few examples see Fig. \ref{fig:maiz}, Fig. 3
from Maiz-Apellaniz 2001). While some (if not most) of them are expected to
dissolve in the near future (see Whitmore 2004 and Bastian et al. 2005 for
``infant mortality'' of young clusters), some have the potential to survive
for a Hubble time to produce a secondary population, similar to the old
globular clusters, but with enhanced metallicity. The {\sl average}
newly-born star cluster does not show signs of tidal truncation (but will
probably develop this during the future interaction with the galactic
gravitational potential), are not (yet) completely relaxed, and the cluster
system as a whole seems to follow a power-law luminosity/mass function
(although this is questioned recently, e.g. de Grijs et al. 2003a,
Goudfrooij et al. 2004,  de Grijs et al 2005b).

\begin{figure}
\begin{center}
\psfig{figure=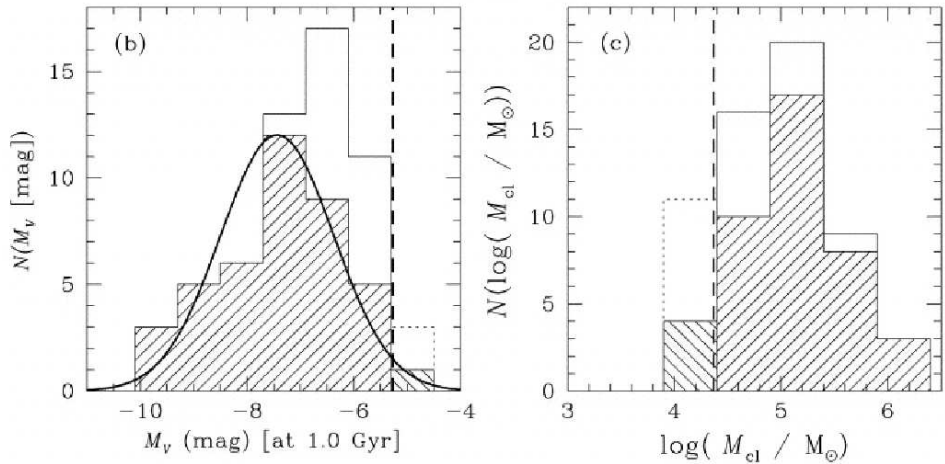,height=2.5in}
\caption{Parameter distributions of clusters in the fossil starburst
region M82-B. The vertical lines indicate selection limit. 
Left panel: Absolute V-band magnitudes, normalised to a 
common cluster age of 1 Gyr. Right panel: Cluster masses. From de Grijs
et al. (2003a)}
\label{fig:m82}
\end{center}
\end{figure}

Therefore, both types of objects seem to be very distinct, though similar
formation mechanisms are expected. The relation between both classes could
be clarified by investigating intermediate-age cluster systems.
Unfortunately, the number of such systems is still small, possibly due to
selection effects. Two well-studied cases are NGC 1316 (Goudfrooij et al.
2004) and the fossil starburst region M82-B (de Grijs et al. 2003a). In
both cases the presence of a turnover (of roughly Gaussian shape) in the
luminosity/mass functions {\sl of the clusters related to the
intermediate-age starburst/merger event} is clearly seen (de Grijs et al.
2003a, Fig. 1 [reproduced here in Fig. \ref{fig:m82}]; Goudfrooij et al.
2004, Fig. 3). However, due to the small number of such systems, the need
to understand the nearby young star cluster formation, evolution and
destruction is still urgent.

\section{Improved photometry for (young) star clusters}

Despite the advances in cluster spectroscopy, the use of integrated,
multi-band photometry is still (and probably will be) the most commonly
used tool to study extragalactic star cluster systems. While CMD analysis
and spectroscopy is always limited to few individual (especially nearby)
clusters, integrated photometry gives, with few exposures, the properties of
the whole cluster sample in the field-of-view.

Though widely used, aperture photometry (and size determination) of
clusters has its caveats which are not yet well studied. For {\sl HST}
observations the most important are the undersampled PSFs, and the
possibility to resolve  star clusters in nearby ($\sim$ 20-25 Mpc)
galaxies. By definition, aperture photometry using finite apertures does
underestimate the total source flux. While this is well studied for
point sources and can be corrected for using aperture corrections (ACs)
(Holtzman 1995), these corrections become increasingly inaccurate with
increasing (apparent) source size.

Therefore, we performed a large-scale study on the size determination
and photometric accuracies for {\sl HST} observations of resolved
clusters. We created artificial clusters using the {\sc BAOlab} package
by S. Larsen (to be obtained from http://www.astro.ku.dk/$\sim$soeren/,
and described in Larsen 1999) as the package to create artificial
clusters most realistic, in conjunction with PSFs and diffusion kernels
produced by {\sc Tiny Tim} (Krist \& Hook 1997).

\begin{figure}
\begin{center}
\includegraphics[angle=270,width=0.48\columnwidth]{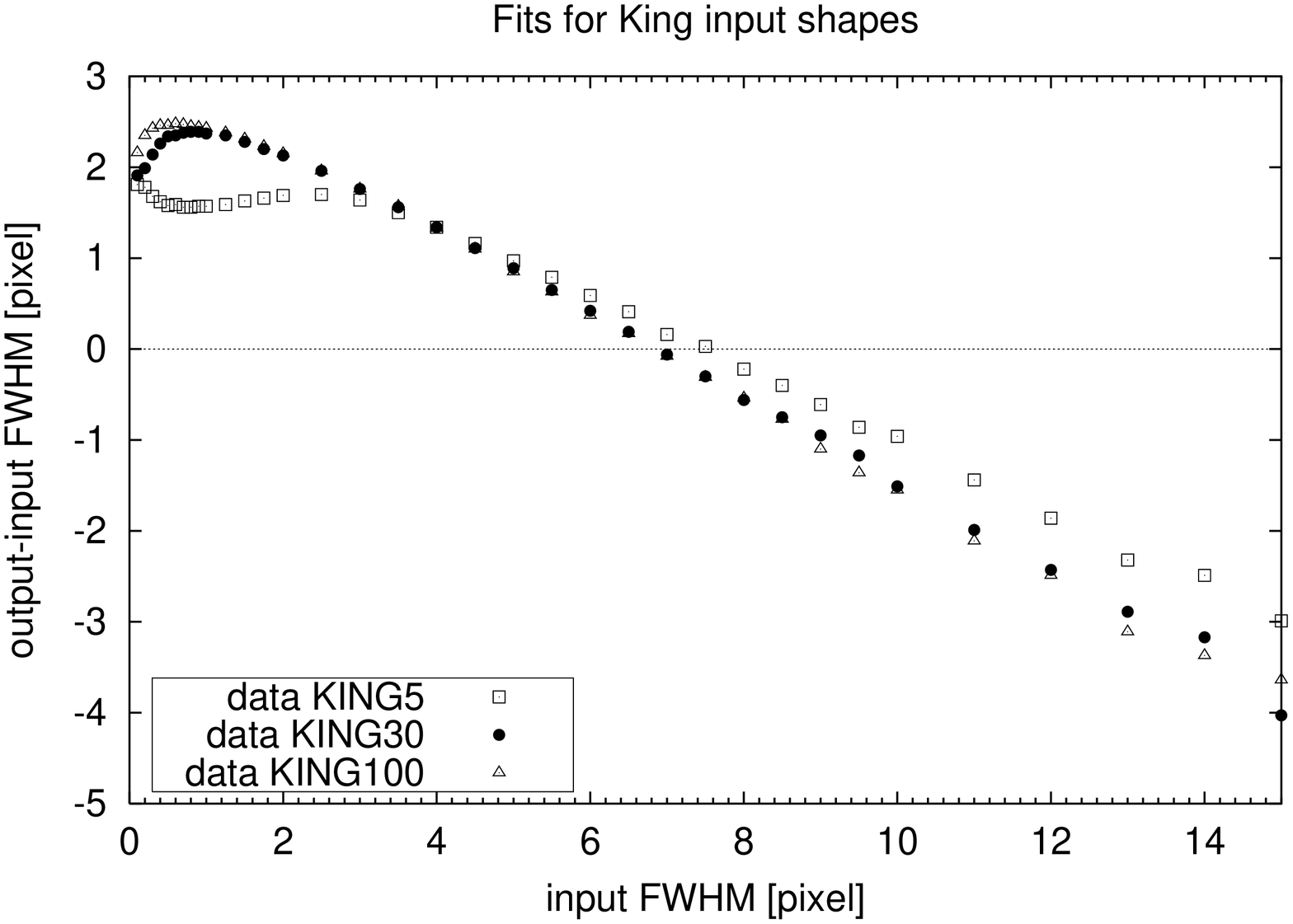}
\includegraphics[angle=270,width=0.48\columnwidth]{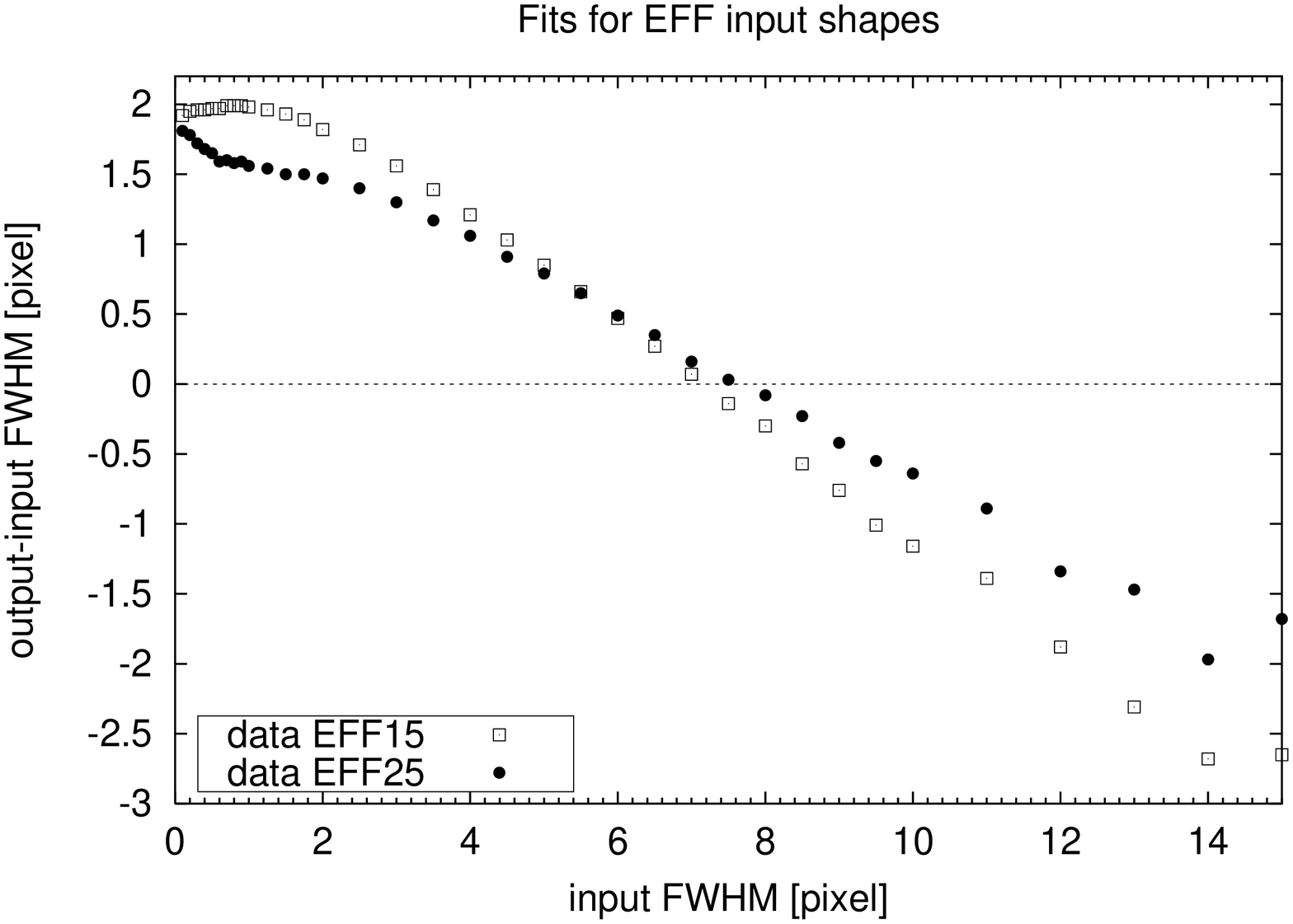}
\caption{Differences between measured (Gaussian) FWHM and input light
profile FWHM (models indicated in legends). Left panel: King (1962)
models with different concentration indices $r_{\rm tidal}/r_{\rm core}$. Right
panel: Elson et al. (1987) models with different power-law slopes.}
\label{fig:sizes1}
\end{center}
\end{figure}

We performed tests with various observational (different {\sl HST}
cameras, chips, filters, sub-pixel shifts), computational (size fitting
radius, aperture annuli) and cluster settings (size, light profile,
brightness, sky background). For each setting we have determined the
relations between input FWHM of the cluster and the FWHM of a fitted
Gaussian (=``measured size'') and parametrised the relations in form of
5$^{th}$ order polynomials. We have chosen to fit a Gaussian for reasons
of general applicability and computational fit stability. Some results
for various input cluster light profiles are shown in Fig.
\ref{fig:sizes1}.

In addition to the cluster sizes we also determined ACs for a number of
typical aperture radii, correcting the source magnitude determined in the
finite aperture to infinite aperture. 

\begin{figure}
\begin{center}
\includegraphics[angle=270,width=0.48\columnwidth]{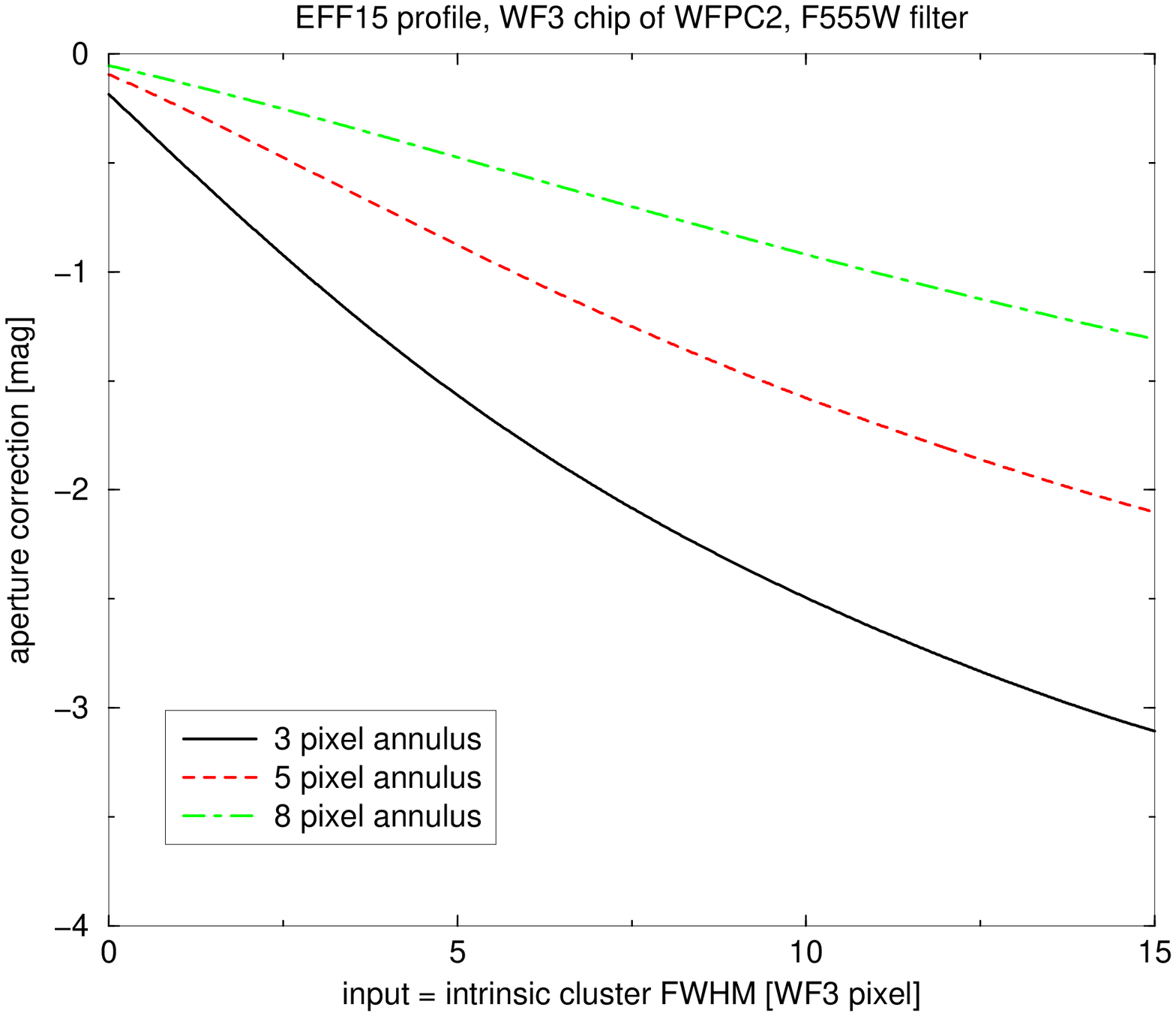}
\includegraphics[angle=270,width=0.48\columnwidth]{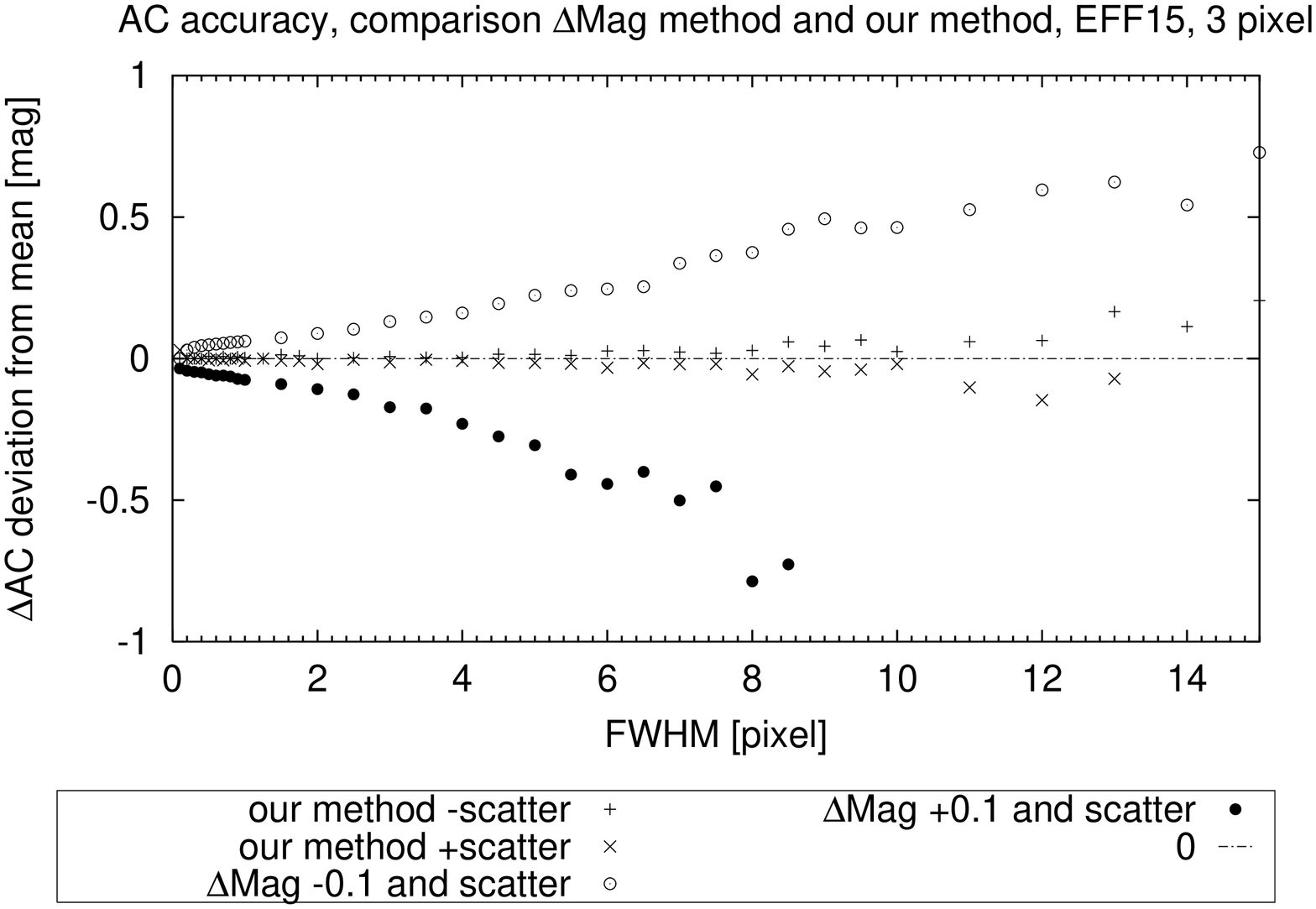}
\caption{Left panel: Theoretical ACs to infinite radius as function of
intrinsic cluster size = cluster FWHM (in WF3 pixel) for 3 different
aperture annuli. Right panel: Comparing the accuracy of AC determination
for our method and the widely used $\Delta$mag method.}
\label{fig:sizes2}
\end{center}
\end{figure}

While fitting the whole cluster image would be the best way to do, the
still most commonly used size measure to date utilises the magnitude
difference in two concentric apertures with different radii centred on
the cluster. Usually radii of 0.5 and 3 pixels are used. Apart from
severe centring problems caused by the use of a 0.5 pixel aperture
radius this method uses only part of the light profile information,
hence a lower accuracy as compared to our method could be expected, and
is proven in Fig. \ref{fig:sizes2}, right panel.

The full set of results (including a ``cookbook'' on how to improve
cluster photometry) will be presented in Anders et al. (2005).

\section{Determining (young) star cluster properties}

Having a set of cluster magnitudes in a number of passbands at hand
enables one to retrieve the physical parameters of this cluster, namely
age, metallicity, extinction within the host galaxy, and mass. This can
be done by comparing the observed Spectral Energy Distributions (SEDs;
the data set containing all magnitudes of a given cluster) with model
predictions, e.g. with {\sc GALEV} models (Schulz et al. 2002, Anders \&
Fritze -- v. Alvensleben 2003, Bicker et al. 2004), GALAXEV = B\&C
models (Bruzual \& Charlot 2003), StarBurst99 (Leitherer et al. 1999) or
PEGASE (Fioc \& Rocca-Volmerange 1997). Many people use these models and
developed their own programs to perform the comparison between models
and observations, usually in a least-square sense. However, these
programs are only briefly explained, mainly in the context of a larger
observational paper. So-far only few large-scale quality assessments or
studies pointing at caveats and strengths of parameter determination by
SED analysis was carried out (from the observational point-of-view e.g. de
Grijs et al. 2003b,c; from a more theoretical point-of-view e.g. de Grijs et
al. 2005a).

\begin{figure}
\begin{center}
\includegraphics[angle=270,width=0.48\columnwidth]{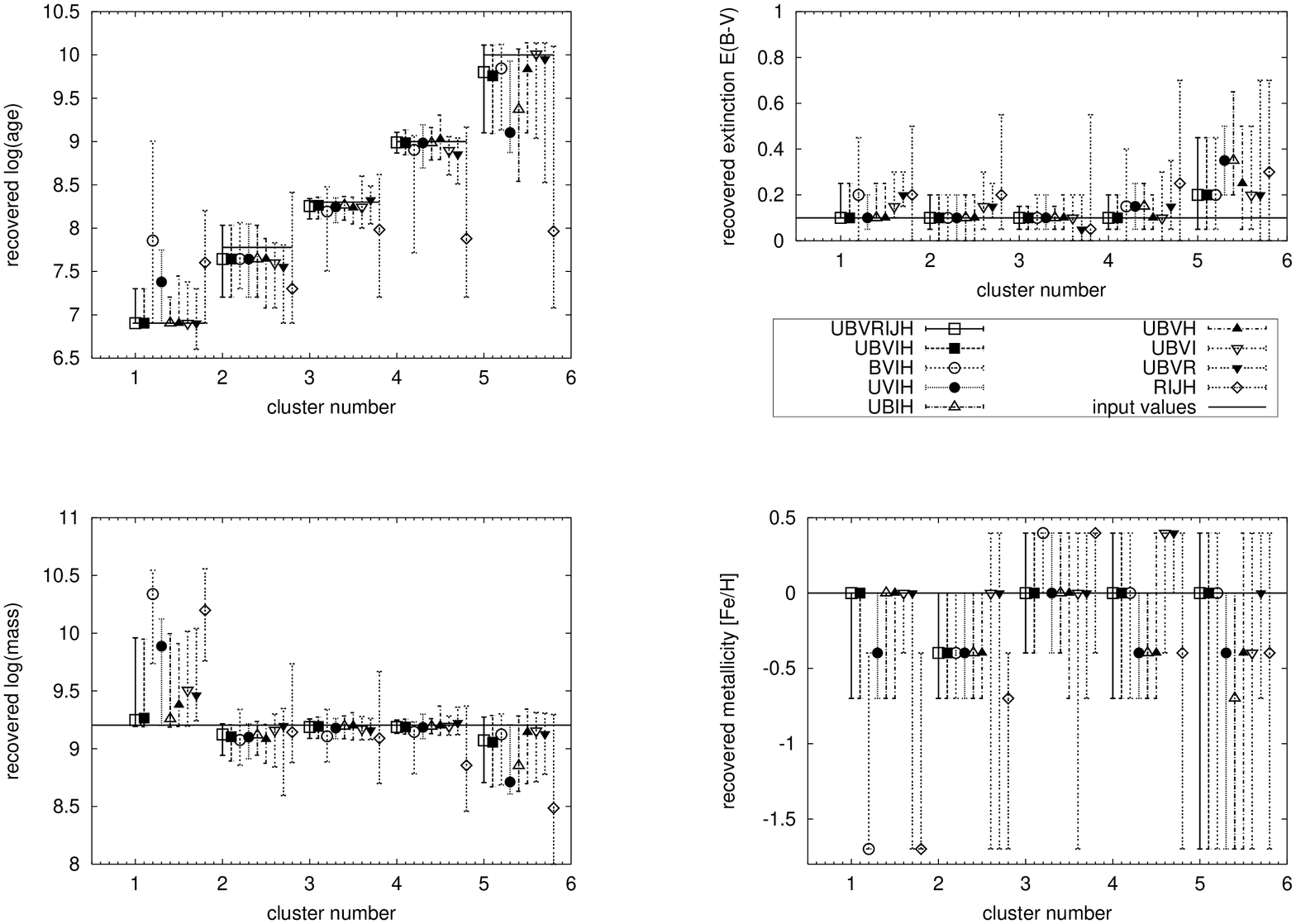}
\includegraphics[angle=270,width=0.48\columnwidth]{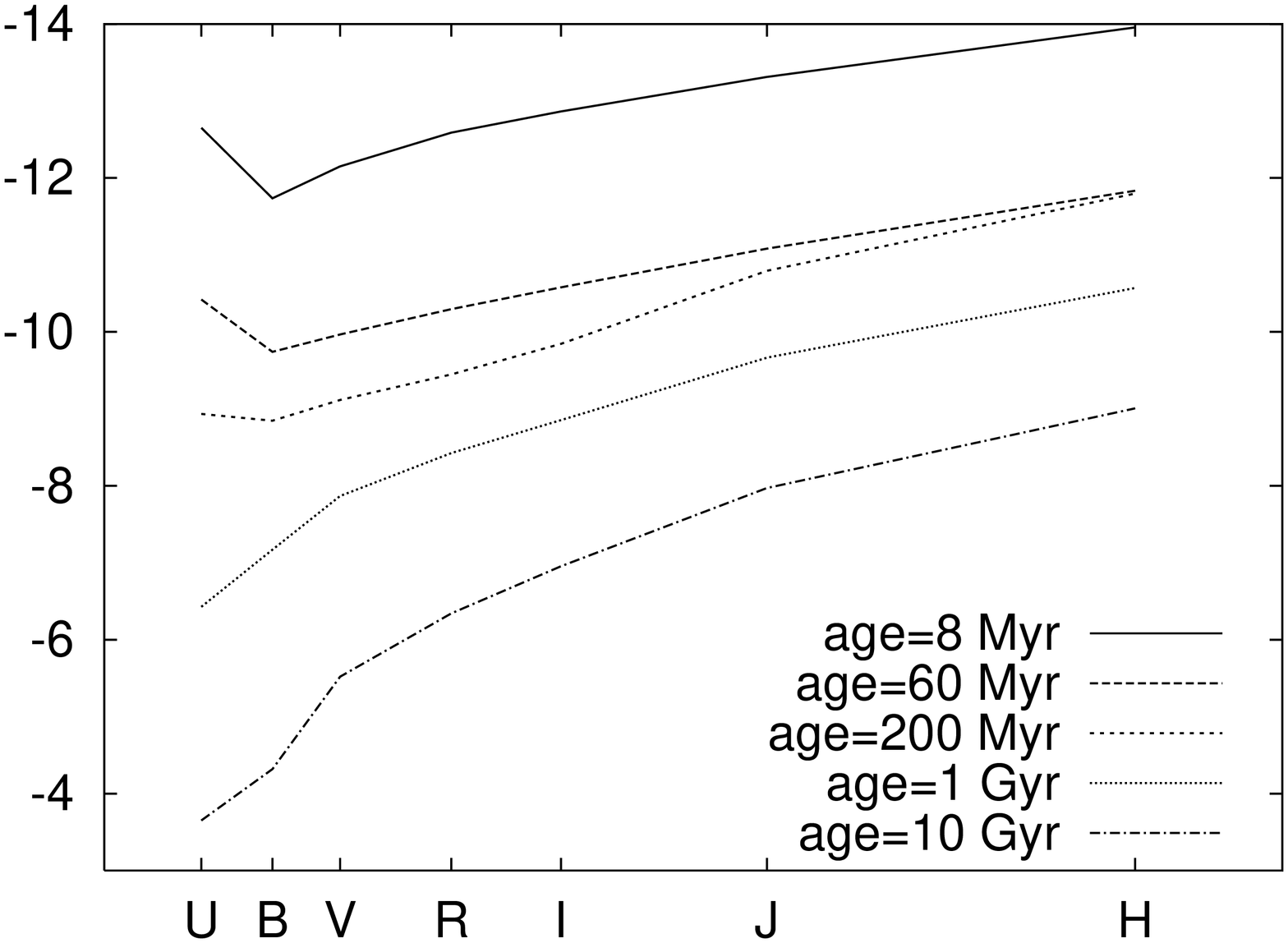}
\caption{Left panels: Comparison of input cluster parameters with retrieved values
for 5 different cluster ages and a number of filter combinations
(indicated in the legend): Age (upper left panel), extinction (upper
right), mass (lower left) and metallicity (lower right). Symbols indicate
the median retrieved values, vertical lines the associated 1$\sigma$
uncertainties, horizontal lines the input values. Right panel: Comparison
of SEDs of different ages.}
\label{fig:param}
\end{center}
\end{figure}

With our {\sc AnalySED} code for SED analysis of clusters we can obtain
the clusters parameters age, metallicity, internal extinction and mass,
and for each parameter the related 1$\sigma$ uncertainties. While
building this code, we carefully evaluated various aspects and caveats
of it (in fact, the general results are valid for all least-square SED
comparison programs). We studied the impact of the choice of filters,
the observational uncertainties, {\sl a priori} assumptions during the
analysis and various cluster parameters (hence, differently shaped
SEDs). For this we have constructed cluster SEDs with known parameters,
added noise to the cluster magnitudes, and then retrieved the cluster
parameters using the {\sc AnalySED} tool, comparing the results with
the input values. In Fig. \ref{fig:param}, left panels, we show the
impact of the filter choice on the parameter determination accuracy, as
an example. As can been seen for the youngest age (8 Myr) the U band
(and to a lesser extent the B band) are most important for accurate
parameter determinations, while for the oldest age (10 Gyr) the B band
(and to a lesser extent the V band) are most important. For filter
combinations strongly biased towards either UV-blue or (even more
pronounced) red-NIR passbands the results show strong deviations from
the input values, misleading the interpretations of the observations.
This confirms the need to carefully choose the filters for
observations, spanning a wavelength range as long as possible. In Fig.
\ref{fig:param}, right panel, we display SEDs of solar metallicity for
5 different ages, from 8 Myr to 10 Gyr to illustrate the changes of
these SEDs with time, and therefore the potential to age-date clusters
based on their SEDs. 

The full study is presented in Anders et al. (2004b).

\section{Application: NGC 1569}

We have applied the AnalySED tool to a data set for clusters in the nearby
dwarf starburst galaxy NGC 1569. We identified a number of 165 clusters
in this galaxy, enlarging the sample of known clusters by roughly a
factor of 3. We find the majority of the clusters to be of low mass,
rather comparable to Galactic open clusters than Galactic globular
clusters. Hence, the vast majority of the clusters are expected to
dissolve in the near future. However, 6 star clusters are more massive
than the median mass of the Galactic globular clusters of 10$^{5.2}
M_\odot \sim 1.6 \times 10^5 M_\odot$. These clusters have the potential
to become globular cluster-like objects in a Hubble time, though they
will lose mass due to dynamical and stellar evolution.

The age distribution confirms the recent burst of star formation inferred
already from CMD analysis and the presence of large {\sc Hii} regions. 

The complete analysis is presented in Anders et al. (2004a).

\newpage
PA and UFvA like to thank the conference organisers for the pleasant and
fruitful conference, as well as for financial support.


\section*{References}
\begin{thebibliography}{99}

\bibitem{2003A&A...401.1063A}  Anders P., \& Fritze-v.~Alvensleben U.,
 2003, A\&A, 401, 1063 

\bibitem{2004MNRAS.347...17A} Anders P., de Grijs  R.,
Fritze-v.~Alvensleben U., \& Bissantz N., 2004a, MNRAS, 347, 17 

\bibitem{2004MNRAS.347..196A} Anders P., Bissantz  N.,
Fritze-v.~Alvensleben U., \& de Grijs R., 2004b, MNRAS, 347, 196 

\bibitem{ich05} Anders P., Gieles M., de Grijs R., MNRAS, {\sl submitted}

\bibitem{nate05} Bastian N., Gieles  M., Lamers H.~J.~G.~L.~M.,
Scheepmaker R.~A., \& de Grijs R., 2005,  A\&A, 431, 905 

\bibitem{2004A&A...413...37B} Bicker J.,  Fritze-v.~Alvensleben U.,
M{\" o}ller C.~S., \& Fricke K.~J., 2004, A\&A, 413, 37 

\bibitem{2003MNRAS.344.1000B} Bruzual G., \&  Charlot S., 2003,
MNRAS, 344, 1000 

\bibitem{1997A&A...326..950F} Fioc M., \&  Rocca-Volmerange B., 1997,
A\&A, 326, 950 

\bibitem{2004ApJ...613L.121G} Goudfrooij P.,  Gilmore D., Whitmore
B.~C., \& Schweizer F., 2004, ApJ, 613, L121 

\bibitem{2003ApJ...583L..17D} de Grijs R., Bastian  N., \& Lamers
H.~J.~G.~L.~M., 2003a, ApJ, 583, L17 

\bibitem{2003MNRAS.342..259D} de Grijs R., 
Fritze-v.~Alvensleben U., Anders P., Gallagher J.~S., Bastian N., 
Taylor V.~A., \& Windhorst R.~A.\ 2003b, MNRAS, 342, 259 

\bibitem{2003MNRAS.343.1285D} de Grijs R., Anders 
P., Bastian N., Lynds R., Lamers H.~J.~G.~L.~M., \& O'Neil E.~J.\ 2003c, 
MNRAS, 343, 1285 
 
\bibitem{2005MNRAS.359..874D} de Grijs R., Anders 
P., Lamers H.~J.~G.~L.~M., Bastian N., Fritze-v.~Alvensleben U., 
Parmentier G., Sharina M.~E., \& Yi S.\ 2005a, MNRAS, 359, 874 
 
\bibitem{richard.gen} de Grijs R., Parmentier G., Lamers H.~J.~G.~L.~M.,
2005b, MNRAS, {\sl submitted}

\bibitem{1995PASP..107..156H} Holtzman J.~A., et  al., 1995, PASP, 107,
156 

\bibitem{1962AJ.....67..471K} King I., 1962, AJ, 67, 471 

\bibitem{krist97} Krist J., Hook R., The Tiny Tim  User's Guide
(Baltimore: STScI)

\bibitem{larsen99} Larsen S.~S., 1999, A\&AS, 
139, 393 
 
\bibitem{1999ApJS..123....3L} Leitherer C., et  al., 1999, ApJS, 123,
3 

\bibitem{2001ApJ...563..151M}  Ma{\'{\i}}z-Apell{\' a}niz J., 2001,
ApJ, 563, 151 

\bibitem{2002A&A...392....1S} Schulz J.,  Fritze-v.~Alvensleben U.,
M{\" o}ller C.~S., \& Fricke K.~J., 2002, A\&A, 392, 1 


\bibitem{2004ASPC..322..419W} Whitmore B.~C.\ 2004, ASP 
Conf.~Ser.~322: The Formation and Evolution of Massive Young Star Clusters, 
322, 419 

\end{thebibliography}
\end{document}